\documentclass[prd,aps,floats,twocolumn,nofootinbib]{revtex4-1}
\usepackage{slashed}
\usepackage{mathtools}
\usepackage{amsfonts}
\usepackage{amssymb}
\usepackage{epsfig}

\begin{document}

\newcommand{\m}[1]{\mathcal{#1}}
\newcommand{\nn}{\nonumber}
\newcommand{\ph}{\phantom}
\newcommand{\eps}{\epsilon}
\newcommand{\be}{\begin{equation}}
\newcommand{\ee}{\end{equation}}
\newcommand{\bea}{\begin{eqnarray}}
\newcommand{\eea}{\end{eqnarray}}
\newtheorem{conj}{Conjecture}


\title{Equivalence of the Chern-Simons state and the Hartle-Hawking and Vilenkin wave functions}
\date{Published 20 Aug 2020}

\author{Jo\~{a}o Magueijo}
\email{j.magueijo@imperial.ac.uk}
\affiliation{Theoretical Physics Group, The Blackett Laboratory, Imperial College, Prince Consort Rd., London, SW7 2BZ, United Kingdom}

\begin{abstract}
We show that the Chern-Simons state when reduced to mini-superspace is the Fourier dual 
of the Hartle-Hawking and Vilenkin wave functions of the Universe. This is to be expected, given that the former and 
latter solve the same constraint equation, written in terms of conjugate
variables (loosely the expansion factor and the Hubble parameter). A number of subtleties in the mapping, related to the contour of integration of the connection, shed light on the issue of boundary conditions in quantum cosmology. 
If we insist on a {\it real} Hubble parameter, then only the Hartle-Hawking wave function can be represented by the Chern-Simons state, with the Hubble parameter covering the whole real line. For the Vilenkin (or tunnelling) wave function the Hubble parameter is restricted to the positive real line (which makes sense, since the state only admits outgoing waves), but the contour also covers the whole negative imaginary axis. 
Hence the state is not admissible if reality conditions are imposed upon the connection. Modifications of the Vilenkin state, requiring 
the addition of source terms to the Hamiltonian constraint, are examined and found to be more palatable. 
In the dual picture the Hartle-Hawking state predicts a uniform distribution for the Hubble parameter
over the whole real line; the modified Vilenkin state a uniform distribution over the positive real line.
\end{abstract}

\maketitle

It is well known that the Chern-Simons state (also called the Kodama state) solves the full, non-perturbative Hamiltonian constraint in the self-dual, or Ashtekar formulation~\cite{jackiw,witten,kodama,lee1,lee2}. The Chern-Simons state is given by:
\bea\label{CS}
\psi(A)&=&{\cal N}\exp {\left(-\frac{3}{2l_P^2\Lambda}Y_{CS}\right)},
\eea
where 
\be
Y_{CS}=\int {\cal L}_{CS}=\int  A^IdA^I+\frac{1}{3}\epsilon_{IJK} A^I A^J A^K
\ee
is the Chern-Simons functional, $A^I$ is the $SU(2)$ Ashtekar self-dual connection (with $I$ its $SU(2)$ indices), $\Lambda$ is the cosmological constant, and 
$l_P^2=8\pi G_N\hbar $.  A number of fair criticisms have been levelled against this state (e.g.~\cite{witten}), namely regarding its non-normalizability, 
CPT violating properties (and consequent impossibility of a positive energy property), and lack of gauge invariance under large 
gauge transformations. All of these criticisms hinge on the fact that the state's phase is not purely imaginary, for example 
proportional to $i\Im Y_{CS}$. If that were the case, then the Lorentzian theory would resemble the Euclidean theory, for which 
these problems evaporate~\cite{randono1,randono2}. We will comment on this issue later in this paper. Suffice it to 
say at this stage that in the minisuperspace approximation the state's phase is {\it always} purely imaginary. 

At this point, we could simply evaluate (\ref{CS}) in minisuperspace without further ado. However, in order to facilitate comparison with
the work of Hartle and Hawking and Vilenkin, we choose an alternative derivation. 
The basis for the Ashtekar formalism is the Einstein-Cartan formulation, upon which a canonical transformation is applied~\cite{thiemann}. 
The reduction of the Einstein-Cartan action to minisuperspace leads to a very simple Hamiltonian system (see e.g.~\cite{alex2,MZ}),
resulting from the action:
\bea\label{ECaction}
S
&=&3\kappa V_c \int dt \bigg(  2a^{2} \dot{b}
+ 2Na\bigg(b^2+k-\frac{\Lambda}{3}a^{2}\bigg) \bigg).
\eea
Here $\kappa=1/(16\pi G_N)$, $a$ is the expansion factor, $b\approx \dot a$ (i.e., on-shell) if there is no torsion, $k=0,\pm 1$ is the spatial curvature and $V_c$ is the comoving volume of the region under study (in most quantum cosmology work $k=1$ and $V_c =2\pi^2$).
Hence the Poisson bracket is:
\bea\label{PB}
\{b,a^2\}&=&\frac{1}{6\kappa V_c}
\eea
and the system reduces to a single constraint (the Hamiltonian constraint) multiplying Lagrange multiplier $N$. 
Quantization of (\ref{PB}) implies:
\bea\label{commutator}
\left[\hat b,\hat{ a^2}\right]&=&\frac{i l_P^2}{3 V_c}
\eea
so that in the $b$ representation:
\be
\hat a^2=-\frac{i l_P^2}{3 V_c}\frac{d}{db}.
\ee
Assuming the ordering implied in (\ref{ECaction}), the quantum Hamiltonian 
constraint equation therefore is:
\be\label{wdweq}
\hat{\cal H}\psi=\bigg(\frac{i \Lambda l_P^2}{9 V_c}\frac{d}{db} + k+b^2 \bigg)\psi=0.
\ee
Its most general solution has the form:
\be\label{kod0}
\psi_{CS}={\cal N} \exp{\left[i\bigg(\frac{ 9 V_c}{\Lambda l_P^2} \left(\frac{b^3}{3}+kb\right)+\phi_0\bigg)\right]}
\ee
where the only ambiguity is in the constant phase $\phi_0$ (which we will set to zero, as it does not 
affect our considerations). 
The real constant $\cal N$ is fixed by the normalization condition: with delta function normalization, as suggested 
in~\cite{randono1}, one has ${\cal N}=1/\sqrt{2\pi}$.
There is no $\pm$ ambiguity in the phase: the plus sign is fixed and will play an important role.

We note that (\ref{kod0}) is nothing but the Chern-Simons state (\ref{CS}) reduced to minisuperspace (as explained, this 
could have been derived directly, right at the start of this paper). Indeed 
for $k=0$ we have $A^I_j=i b \delta^I_j$, leading to (\ref{kod0}) trivially. 
The calculation is more involved for $k\neq 0$ (see~\cite{MZ,MSZ}), but the conclusion remains true. 
This is hardly surprising, since equation (\ref{wdweq}) is nothing but a minisuperspace reduction of the 
full Hamiltonian constraint with an appropriate ordering.  
Note that the fluxes conjugate to $A^I_a$ are the densitized inverse triads $E^a_I$ 
which in minisuperspace become $E^a_I=a^2\delta^a_I$, in agreement with (\ref{PB}). We will say more about this later, but we stress 
that from this perspective it is clear that the base variable for discussing quantum cosmology in the metric representation should 
be not the expansion factor, $a$, but its square, $a^2$. This apparently innocent remark has many a radical implication. 

We now move on to the main point of this paper, and the reason for our alternative derivation of
(\ref{kod0}).  Our minisuperspace Hamiltonian constraint equation (\ref{wdweq}) is nothing but 
the standard Wheeler-DeWitt equation in the complementary 
representation implied by  commutator (\ref{commutator}). 
Had we chosen the metric (or, rather, the $a^2$) representation, then:
\be
\hat b=\frac{i l_P^2}{3 V_c}\frac{d}{d(a^2)}
\ee
and the Hamiltonian constraint equation would have read:
\be\label{wdweq-a}
\left[\frac{d^2}{da^2}-\frac{1}{a}\frac{d}{da}-U(a)\right]\psi=0,
\ee
with 
\be\label{potential}
U(a)=4\left(\frac{3V_c}{l_P^2}\right)^2 a^2\left(k-\frac{\Lambda}{3}a^2\right).
\ee
This is just the usual Wheeler-DeWitt equation with a specific ordering. 
We use the excellent  review by Vilenkin~\cite{Vilenkin} as the gold standard. 
Setting $k=1$, $V_c=2\pi^2$, and choosing the ordering parameter (as defined in~\cite{Vilenkin}) $\alpha=-1$, we find that  indeed there is agreement\footnote{To bridge notation notice that $6\kappa V_c=1/2\sigma^2$ as defined in~\cite{Vilenkin}. Also note that $\Lambda$ is defined with an extra factor of 1/3 there.}. This is not surprising, since the Einstein-Cartan action reduces to the Einstein-Hilbert action
if there is no torsion. The solutions of this equation include the 
Hartle and Hawking~\cite{HH,Vilenkin} and the Vilenkin or tunnelling wave functions~\cite{vil0,Vilenkin}, 
depending on which boundary conditions one adds to this equation.

What can, therefore,  be the relation between the Hartle-Hawking and Vilenkin wave functions, on the one hand, and the Chern-Simons state,
on the other? Obviously, {\it in some sense}, the two have to be related by a Fourier transform, since they solve the same quantum equation in terms of  complementary variables. The Fourier transform inferred from (\ref{commutator}) is:
\bea\label{FT}
\psi_{a^2}(a^2)&=&\frac{3V_c}{l_P^2}\int \frac{db }{\sqrt{2\pi}} e^{-i\frac{3V_c}{\l_P^2}a^2 b}\psi_b(b).
\eea
But at once we notice an oddity. The Wheeler-DeWitt equation in the metric representation is second order
(allowing two linearly independent solutions: the Hartle-Hawking and Vilenkin wave functions), whereas in the $b$ representation it is first order,
so that the Chern-Simons wave function is essentially unique up to an irrelevant phase and normalization constant. 
This points to an ambiguity in the Fourier transform, capable of incorporating this disparity
in degrees of freedom. Resolving the matter will explain how the
 Chern-Simons state can be dual to  {\it both} the Hartle-Hawking and the Vilenkin proposals. 

The simplest way to unveil the detailed map
is to examine concrete
solutions.
%
In the $a^2$ representation these are Airy-type functions~\cite{vil0},  specifically:
\be
\psi_V\propto {\rm Ai}(-z)+i{\rm Bi}(-z) 
\ee
for Vilenkin boundary conditions, and
\be
\psi_H\propto {\rm Ai}(-z),
\ee
for Hartle-Hawking boundary conditions, with:
\be\label{zexp}
z=-\left(\frac{9V_c}{\Lambda l_P^2}\right)^{2/3}
\left(k-\frac{\Lambda a^2}{3}\right).
\ee
We can now appeal to well-known results in the theory of Airy functions~\cite{Airy-book,Abra-book}
familiar in optics and quantum optics.  These special functions
have integral representation
\be\label{intrep}
\phi (z)=\frac{1}{2\pi}\int  e^{i\left(\frac{t^3}{3}+zt\right)}dt 
\ee
where $\phi$ can be Ai,  Bi or a combination thereof depending on the choice of contour over which the $t$ integration is
undertaken. {\it It is a central result of this paper} that inserting (\ref{kod0}) (the Chern-Simons state) into (\ref{FT}) (the proposed Fourier transform) 
leads precisely to integral (\ref{intrep}) with replacements (\ref{zexp})
and:
\be\label{texp}
t=\left(\frac{9V_c}{\Lambda l_P^2}\right)^{1/3}b.
\ee
Hence the Chern-Simons wave function is indeed the Fourier dual of the Hartle-Hawking and Vilenkin wave functions, 
with the choice of range for the connection $b$ (or of contour for the  integral (\ref{intrep})) dictating which of the two functions is represented.

\begin{figure}
	\center
	\epsfig{file=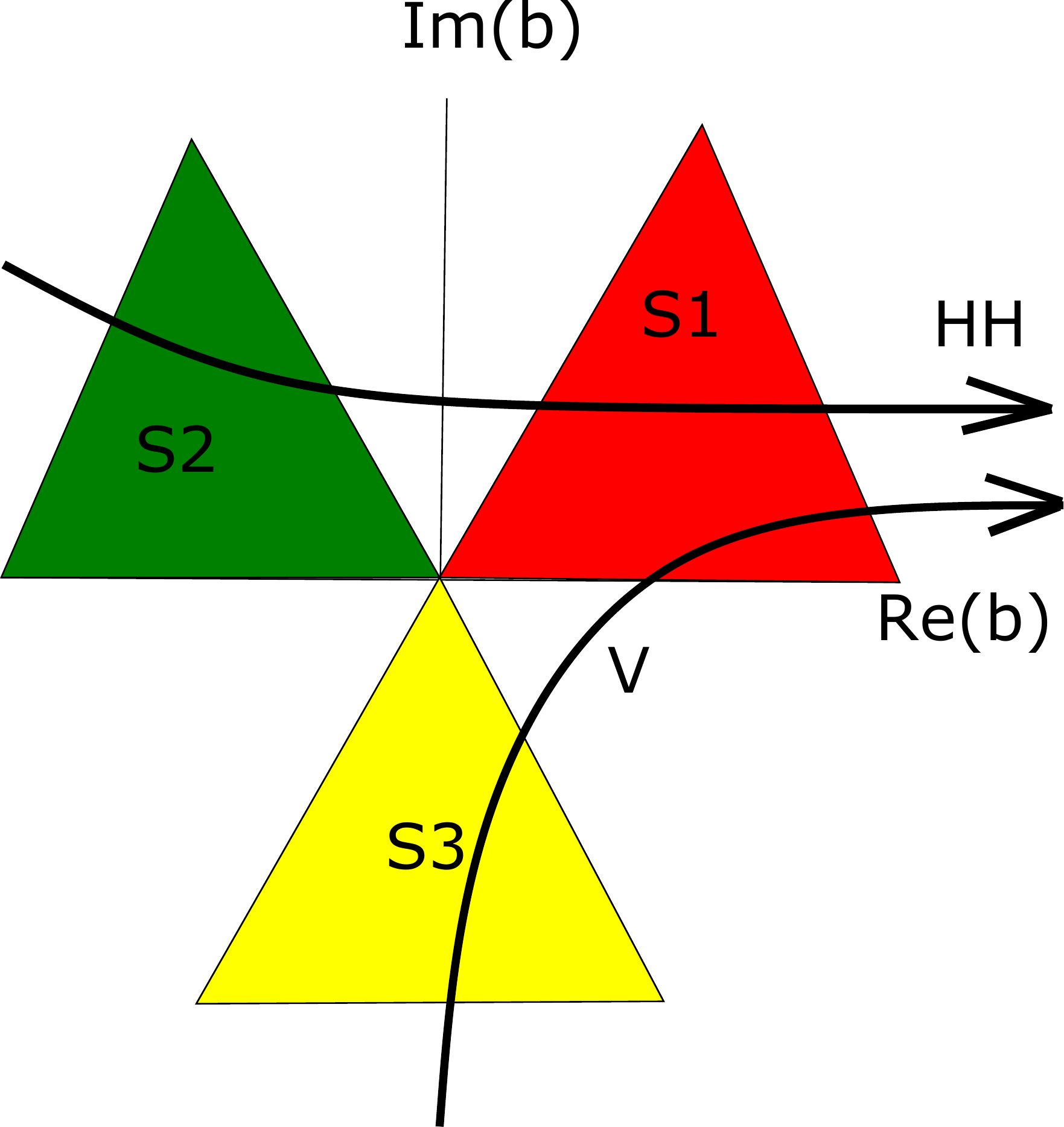,width=8 cm}
	\caption{The 3 sectors S1 (red), S2 (green) and S3 (yellow) where the contour of the integral representation must start and
	finish (at infinity). The upper line illustrates a choice of contour leading to the Hartle-Hawking wave function; the lower line a contour leading to the Vilenkin wave function. }
	\label{fig1}
\end{figure}

This choice can be identified from standard results~\cite{Airy-book,Abra-book}.
The integration contour must start and finish at complex infinity 
within one of the 3 sectors:
\bea\label{sectors}
{\rm S1:}\;\;\; \quad 0<&\arg (t)&<\frac{\pi}{3}\nn\\
{\rm S2:}\quad \frac{2\pi}{3}<&\arg (t)&< \pi\nn\\
{\rm S3:}\quad \frac{5\pi}{3} <&\arg (t)&<\frac{7\pi}{3},
\eea
as depicted in Fig.~\ref{fig1}. 
This is because (\ref{intrep}) only solves the Airy equation following an
integration by parts, producing a boundary term that requires the integrand to vanish at the endpoints. 
Any contour starting (at infinity) in sector S2 and finishing (at infinity) in sector S1 produces the Hartle-Hawking function. 
Instead, any contour starting in S3 and finishing in S1 produces the Vilenkin wave function. These are the only independent
possibilities\footnote{Obviously an ``anti-Vilenkin''  wave function containing only incoming waves could be built with a contour starting in S3 and finishing in S2, but then the Hartle-Hawking contour would be the linear combination of this contour and Vilenkin's. Any contour starting and finishing in the same sector does not enclose any pole, and therefore leads to zero.}. Examples of such contours are drawn in Fig.~\ref{fig1}. 


\begin{figure}
	\center
	\epsfig{file=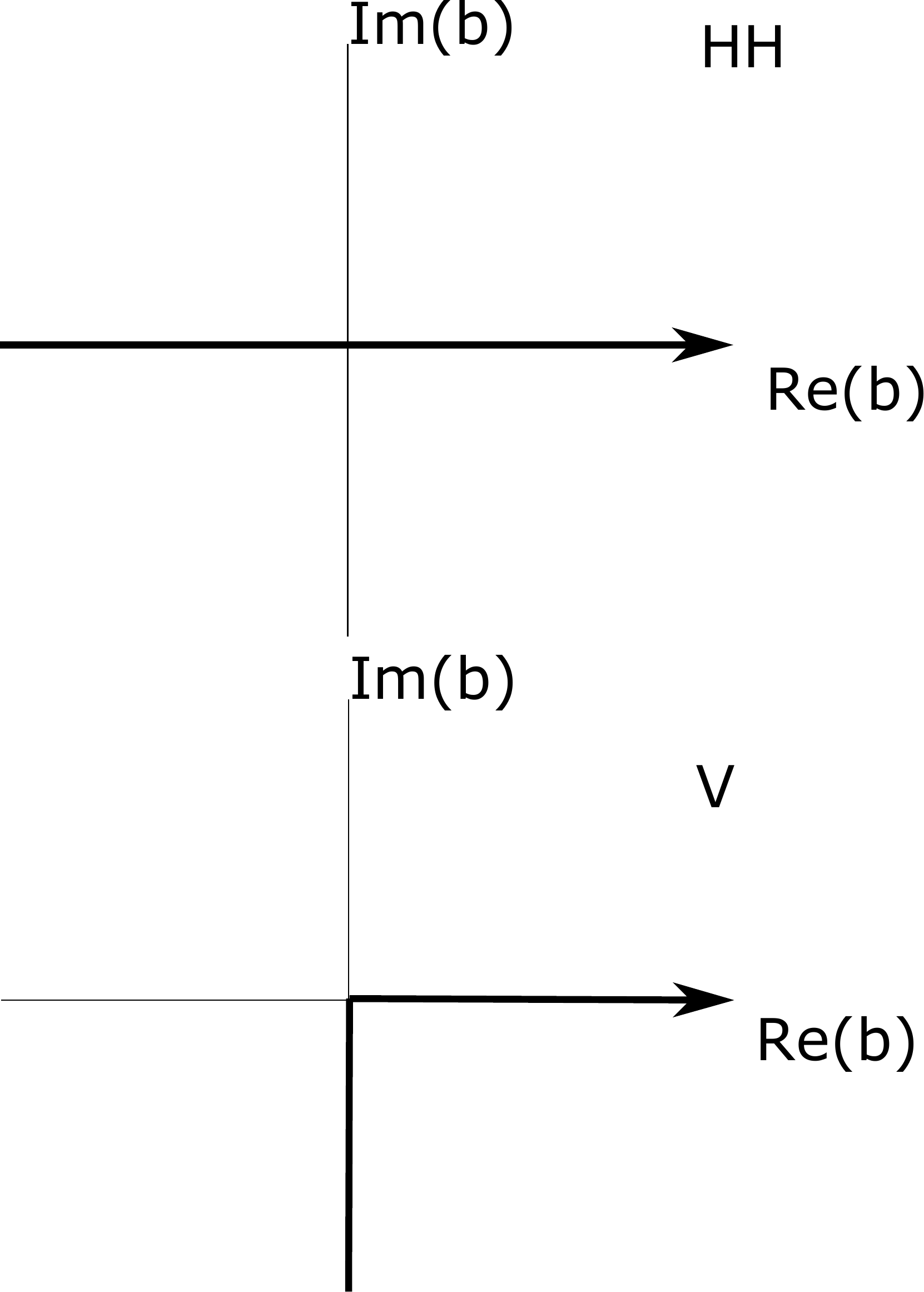,width=8 cm}
	\caption{The simplest choice contours of integration for the connection/Hubble parameter $b$ (classically identical to $\dot a$). 
	The top contour leads to the Hartle and Hanwking wave function and is the real line. The bottom one leads to the Vilenkin/tunneling one, and is made up of the negative imaginary line and the positive real line.} 
	\label{fig2}
\end{figure}

Two further qualifications are in order. Firstly, the inequalities defining sectors (\ref{sectors}) 
may be non-strict (i.e. include equalities) if we accept 
an extended sense of convergence~\cite{Airy-book,hardy}. This will include delta-function normalization of the Chern-Simons state, as we shall see. Secondly, even though the contours just described are completely generic, two particular choices stand out. If we insist on $b$ being real, then only the Hartle-Hawking wave function can be dual to the Chern-Simons state. The integral in (\ref{intrep}) is then to be seen as over the real line, containing both expanding and contracting Universes (see Fig.~\ref{fig2}). Should we required strict convergence (non-delta function normalization) we can shift the contour by:
\be\label{epsshift}
b\rightarrow b+i\epsilon 
\ee
and then let $\epsilon\rightarrow 0$, as is common in some QFT integrations. 
In contrast, the Vilenkin wave function requires any integration over real $b$ to extend at most over the positive axis only. This makes sense,
since the wave function is restricted to representing outgoing waves after tunnelling. However, the integral cannot start at zero and then
follow the positive real axis. Such integral represents Scorer's functions $\rm Gi$ and $\rm Hi$ and they solve a
different differential equation~\cite{Abra-book} (i.e. not the Airy/Wheeler-DeWitt equation):
\be\label{scorer}
\psi''+z\psi=\frac{1}{\pi}.
\ee
Hence, to obtain the Vilenkin wave function, we must allow for imaginary values of $b$, for example, starting along the negative imaginary 
axis, then swerving into the positive real axis at the origin (see Fig.~\ref{fig2}). The implications will be studied below.

Although all of this is standard mathematics, alternative derivations may be found which 
make contact with familiar results in quantum cosmology.
For example, the WKB approximations (often used in quantum cosmology~\cite{Vilenkin} both in the non-oscillatory regime, under the``barrier'' of $U(a)$, and in the oscillatory regime at large $a$) can be recovered from a stationary phase 
approximation to  (\ref{intrep}). The derivation is instructive. The integrand in (\ref{intrep}) may be written as $e^{i{\cal S}}$, not to be confused with the Euclidean path integral\footnote{We stress that 
${\cal S}$ is a function of $b$, so we are  not integrating over the metric, but over the connection. Also ${\cal S}$ is made up of the Chern-Simons functional, rather than the Hamiltonian, so it is not the euclidean action. This is not the usual Hawking integral in quantum cosmology.}.  Unwrapping the integral in its full glory it reads:
\bea
\psi_{a^2}(a^2)&\propto&\int \frac{db }{2\pi} \exp{\left[\frac{ 9 i V_c}{\Lambda l_P^2} \left(\frac{b^3}{3}+kb
-\frac{\Lambda ba^2}{3}\right)\right]}
\eea
so that:
\be
\frac{\partial {\cal S}}{\partial t}\propto \frac{\partial {\cal S}}{\partial b}\propto {\cal H}
=b^2+k-\frac{\Lambda a^2}{3}.
\ee
Hence the stationary points of phase ${\cal S}$ (containing the Chern-Simons functional, not the Hamiltonian) are the solutions to the 
classical Hamiltonian constraint ${\cal H}\approx 0$, given by:
\be
b_{\pm}=\pm\sqrt{\frac{\Lambda a^2}{3}-k}
\ee
(or the equivalent expression in terms of $t$ and $z$, according to (\ref{texp}) and (\ref{zexp})). By taking the Taylor expansion to second order around these points:
\be
{\cal S}_\pm=-\frac{2}{3}t_\pm^3 + t_\pm (t-t_\pm)^2
\ee
we find that the integral (\ref{intrep}) can be carried out (see the relevant Appendix in~\cite{Airy-book} for details). 
It leads to the WKB expressions for the Hartle-Hawking wave function if
both ${\cal S}_\pm$ are included; to the Vilenkin wave function if only ${\cal S}_+$ is selected. We found this to be the simplest way to make contact with these well-trodden territories\footnote{It is tempting to change variables from $t$ to ${\cal S}$ itself in (\ref{intrep}), knowing that ${\cal H}$ would then appear in the denominator of the transformed integral. Unfortunately the two poles thus generated in fact form a branch cut (since ${\cal S}$ is multivalued between them), and so do not fall within the remit of the residue theorem. This prevents a nice connection with the Feynman, advanced and retarded propagators, as far as we can see.}.

So far our equivalence is purely formal, but what can all of this mean? We feel that a deep connection between these two hitherto
separate fields must exist. In the final part of this paper we content ourselves with picking the lowest-hanging fruit, hoping to motivate further work.

First of all, we learn an important lesson about the Kodama/Chern-Simons wave function: that this function, by itself, does not fix a 
quantum state. To turn it into a quantum state one must specify the range (or contour) of the connection, in lieu of what are standard boundary conditions in the dual metric representation. Thus, we should distinguish between the Chern-Simons wave function defined for  $b\in D_1=\mathbb{R}= (-\infty,\infty)$ and  for  $b\in D_2=(-i\infty,0)\cup (0,\infty)$.

Once this is recognized, however, it makes little sense to distinguish between the Chern-Simons
{\it state} (i.e. function and its domain) on the one hand,  and the Hartle-Hawking or the Vilenkin wave functions, on the other. These two``hands" are  {\it the same quantum state} expressed  in different representations. Only
$|\psi_V\rangle$ exists in Hilbert space, with:
\bea
\psi_V(a^2)&=&\langle a^2|\psi_V\rangle \\
\psi_{CS}(b; b\in D_2)&=&\langle b|\psi_V\rangle. 
\eea
and likewise for $|\psi_{HH}\rangle$. Having understood this simple but important fact we may now benefit from the cross-pollination 
resulting from examining the same quantum state from the complementary perspectives of conjugate variables.

Foremost, we find the issue of reality conditions in the Ashtekar formulation~\cite{thiemann}. At face value, these imply a rejection of the Vilenkin state. The reality conditions require the reality/hermiticity of $E^a_I$ and that the anti-self dual connection $\bar A^I_a$ be the complex/hermitian conjugate  of the self-dual connection $A^I_a$. Thus, in minisuperspace the reality conditions imply that $a^2$ and $b$ must be real. This disqualifies the Vilenkin wave function, due to its compulsory foray into the negative imaginary axis of $b$, 
but it is possible that a less strict interpretation of the reality conditions might change this conclusion. 
We stress that the Vilenkin state's forced inclusion of the imaginary axis for $b$ (the Hubble parameter) has nothing to do with the tunnelling property of the wave function, and its having support under the barrier of $U(a^2)$. The Hartle-Hawking wave function also has support in this classically forbidden region, and yet its $b$ dual can live on the  real line only.

Curiously, in the reverse direction, the reality conditions only require that $a^2$ be real: they do not require that it be positive. 
As already pointed out, the fluxes of $A^I_a$ are the densitized inverse triads $E^a_I$ and these are proportional to $a^2$ in minisuperspace. 
This is also the canonical variable conjugate to $b$, and it explains why all the 
wave functions are functions of $a^2$ alone. Therefore $a^2\in (-\infty,\infty)$ is natural, coming from the connection perspective; indeed this is needed to render
(\ref{FT}) invertible and the basis in $a^2$ complete. Strangely, the reality conditions imply that we need to consider
Euclidean regions for the FRW metric. This has consequence for the normalization of the various wave functions, a matter we now 
turn to.



The Chern-Simons state, under the guise of the Kodama state, has been much maligned on the grounds of its non-normalizability, among other perceived deficiencies (e.g.~\cite{witten}). 
As already pointed out at the start of this paper, these crimes vanish for the state's Euclidean formulation~\cite{randono1,randono2}, where the state becomes a pure phase (i.e. its exponent is imaginary). As we have seen here the same happens in the Lorentzian 
theory in minisuperspace. Elsewhere~\cite{MSZ,MSZ1}
we will show that it is possible to mimic the minisuperspace treatment starting from the Einstein-Cartan action (\ref{ECaction})
in the full theory. This leads to the modified state:
\bea\label{kodrealth}
\psi_{CS}& =&{\cal N}'\exp {\left(-\frac{3 i}{l_P^2\Lambda} \Im Y_{CS}\right)}.
\eea
With the imaginary phase property of the wave function assured, the state is as normalizable as a plane wave extending over 
the whole state, i.e. it is delta-function normalizable, belonging to a rigged Hilbert space. 
This assumes, of course, the reality of $b$.

\begin{figure}
	\center
	\epsfig{file=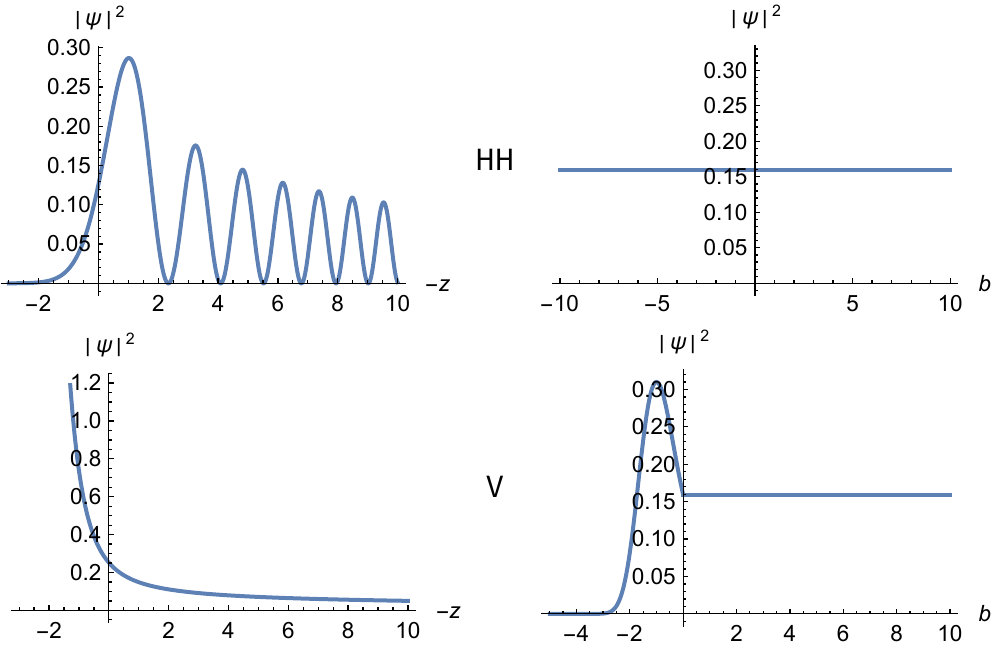,width=8 cm}
	\caption{The probability density as a function of $-z$ (i.e. in the $a^2$ representation) and as a function of $b$, 
	for Hartle-Hawking and Vilenkin quantum states. In the bottom left plot the $b$ axis is real for positive arguments, imaginary for negative
	arguments, as explained in the text.
	} 
	\label{fig3}
\end{figure}

A probabilistic interpretation may now be attempted. For the Hartle-Hawking state, the wave function in $b$ space is a pure phase over the whole of its domain, so the prediction is a uniform distribution in $b$ over the real line. This is to be interpreted in the same way as the probability distribution in space for a plane wave extending over the whole space. It can be regulated with a UV cut-off in $|b|$, for example, or else with prescription (\ref{epsshift}). Such a uniform distribution is the flip-side of the the distribution of $a^2$ implied by the Hartle-Hawking 
wave function (i.e. $
P_{HH}(a^2)\propto Ai^2(-z)
$, see Fig.~\ref{fig3}).
This is of course not uniform, indeed in the classically allowed region the wave function is a standing wave, so the probability 
is modulated by oscillations.
The fact that this is not strictly convergent  as $a \rightarrow \infty$ reflects the same issues found in the 
$b$ representation: that the state is only delta-function normalizable. In terms of the $a^2$ representation this means that:
\be
\int_{-\infty}^\infty  dz\psi_{HH}^\star (z+x)\psi_{HH}(z+y)=\delta(x-y).
\ee
We can still make sense of relative probabilities over the whole $a^2\in (-\infty,\infty)$. 
As $a^2\rightarrow-\infty$ the probability dies down exponentially. 

The probability distribution for the Vilenkin state in $b$ space is more difficult to interpret, given that $b$ must abandon the real
line. Naively, the state predicts a uniform distribution in $b$ over the positive real line. 
Over the negative imaginary line, written as $b=i\Im(b)$, the prediction is:
\be
P_V(\Im b )=\frac{1}{2\pi}\exp{\left[\frac{ 18  V_c}{\Lambda l_P^2} \left(\frac{\Im b^3}{3}- k \Im b\right)\right]}
\ee
rising to a peak at $b=-i$ (for $k=1$), then falling off exponentially to zero, as $b\rightarrow -i\infty$ (see Fig.~\ref{fig3}).
How these conclusions map into $a^2$ space is less obvious. 
Note that in $a^2$ space the Vilenkin wave function does not exhibit the same modulations as the Hartle-Hawking wave function in the classically allowed regime,
since it is a travelling wave. 
More importantly, given that $a^2$ (seen as a dual to $b$) should extend to minus infinity, the state appears problematic.
As $a^2\rightarrow -\infty $ the Vilenkin state diverges exponentially (due to the ${\rm Bi}$ function). Hence the regulating procedure analogous to that proposed for Hartle-Hawking should {\it not} exist.

Naturally the tunnelling state can be retouched to make it more palatable. It was suggested (e.g.~\cite{Vilenkin}) that the state is only non-zero for $a>0$, in which case it solves a modified Wheeler-DeWitt equation, with a delta-Dirac source term. Vilenkin's state is then
a Green's function of the Wheeler-DeWitt operator. This is a different wave function and quantum state, which we will label V1.
Although it was obtained with more sophisticated methods
(e.g. the path integral formalism) a pedestrian derivation follows from writing:
\be
\psi_{V1}(a^2)=\langle a^2|\psi_{V1}\rangle=\psi_V(a^2)\Theta(a^2).
\ee
Insertion into (\ref{wdweq-a})  generates a source term in $\delta(a^2)$. Had we dressed $\psi_V(a^2)$ with
$\Theta(a)$ a source term proportional to in $\delta(a^2)$ would also have been obtained. 

We stress that this wave function represents a state different from $|\psi_V\rangle$. It solves a different equation. 
As in the case of the range of $b$ and the Chern-Simons function, the range of $a^2$ now becomes as relevant in defining the state
as the function itself. 
Its dual representation
$\psi_{V1}(b)= \langle b|\psi_{V2}\rangle$ no longer is
the Chern-Simons wave function, subject to whatever contour. A source term proportional to $\delta(a^2)$
in (\ref{wdweq-a}) translates into a constant source term in the $b$ dual representation, 
Eq.~(\ref{wdweq}). The Chern-Simons wave function is not a solution. Elsewhere we will study the modified
wave function in the connection representation associated with this state. 

\begin{figure}
	\center
	\epsfig{file=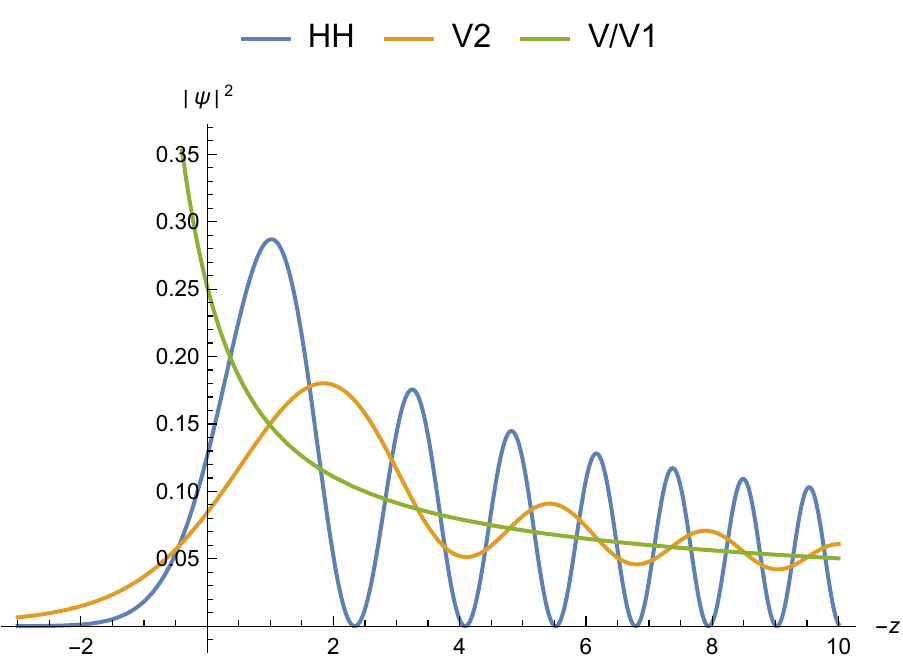,width=8 cm}
	\caption{The probability density $|\psi(z)|^2$ as a function of $-z$ for the Hartle-Hawking state and the various Vilenkin state interpretations.  Note that V1 is just V subject to $a^2>0$ (or $a>0$, if required). The state V2 is a dual of the Chern-Simons state which represents expanding Universes only, and satisfies the reality conditions. It falls off exponentially for $a^2\rightarrow -\infty$ just like the Hartle-Hawking state. 
	} 
	\label{fig4}
\end{figure}

Here, instead, we will do something simpler. 
Once we accept the introduction of delta-function sources in the Wheeler-DeWitt equation as justified means 
to an end (that of imposing desirable domain truncations) there is no reason not to do it in the $b$ representation. 
We therefore backtrack to the point in this paper (around Eq.~(\ref{scorer})) where we dismissed the possibility
of starting the $b$ contour at the origin, following the positive real axis only.
Such a modified tunnelling state (which we label $|\psi_{V2}\rangle$) is defined by
\be
\psi_{V2}(b)=\langle b|\psi_{V2}\rangle= \psi_V(b)\Theta(b)
\ee
that is, 
\be
\langle b|\psi_{V2}\rangle=\psi_{CS}(b; b\in D_3)
\ee
or the Chern-Simons wave function  
with $b\in D_3= (0,\infty)$.  Insertion into Eq.~(\ref{wdweq}) leads to a 
source proportional to $\delta(b)$. This would Fourier transform into a constant source term in  Eq.~(\ref{wdweq-a}):
the source term associated with $|\psi_{V2}\rangle $ is the Fourier dual of that for  $|\psi_{V1}\rangle$. 
This constant term in the Wheeler-DeWitt equation in the $a^2$ representation is nothing but the RHS of Scorer's equation (\ref{scorer}).
We thus arrive at the tunnelling wave function in $a^2$ space:
\be\label{V2a2}
\psi_{V2}(a^2)=\langle a^2|\psi_{V2}\rangle\propto {\rm Ai} (-z)+ i {\rm  Gi} (-z).
\ee
We plot its associated probability density, compared to other proposals, in Fig.~\ref{fig4}. The probability of $b$ is uniform
over the positive real line.

The state $|\psi_{V2}\rangle$ is very interesting. It shares with other Vilenkin-like proposals the feature that it contains only outgoing waves
(in the sense that its transform only contains $b>0$, i.e. expanding Universes, with $\dot a>0$).
Wave function (\ref{V2a2}) is the dual of the Chern-Simons wave function defined over a contour that complies with the 
reality conditions. Indeed the offensive contribution (the integral (\ref{intrep}) over the imaginary negative axis) has form:
\be
\phi_{IM} (z)=\frac{i}{\pi}\int_{-\infty}^0  e^{\left(\frac{\bar t^3}{3}-z\bar t\right)}d\bar t =i{\rm Hi}(-z),
\ee
with $\bar t=\Im (t)$. But since ${\rm Bi}={\rm Gi}+{\rm Hi}$, we have:
\be
\psi_{V2}(a^2)=\psi_{V}(a^2)-\psi_{IM}(a^2),
\ee
complying with the reality conditions. 
This state is also well behaved as $a^2\rightarrow -\infty$, just like
the Hartle-Hawking state.  Is this the best of both worlds?

To conclude, perhaps the most radical implication of the exploration of dual pictures pursued in this paper is the damning of ``creation of the Universe out of nothing". ``Nothing" here is $a=0$, but coming from the canonical perspective which gives primacy to the connection, the natural 
dual variable is $a^2$, the densitized metric. The question then is not nucleation from nothing ($a=0$, excising $a<0$ ), but whether or not to include the Euclidean section ($a^2<0$). From the connection perspective there is no reason not to consider
$a^2\in(-\infty,\infty)$. The relevant issue is therefore, what is the probability for a Lorentzian Universe, $P_L$? 
For the Hartle-Hawking, V1 and V2 states it is 1. For the unexpurgated Vilenkin state it is zero. 

The point $a=0$ is unexceptional. Also, all our results are functions of $z$ alone (defined in (\ref{zexp})), so they 
apply equally well to non-spherical Universes ($k=0,-1$). For $k=0,-1$ we can consider 
topologically non-trivial versions with finite $V_c$ and integrate over the whole space; or 
we can consider the quantum mechanics of a given finite comoving region. Whatever the case, 
the results are essentially the same.  Different choices of $k$ (as well as $\Lambda>0$ and $V_c$) merely shift the value of $z$ where
Euclidean gives way to Lorentzian spaces, but the results found are generic.
 For  $\Lambda<0$ the relation between the sign of $a^2$ and that of $z$ reverses, so our conclusions
reverse. Negative Lambda seems to favour Euclidean Universes.

I would like to thank Alex Vilenkin for many patient interchanges, and Simone Speziale and 
Tom Zlosnik for collaboration on a project of which this paper is a minor offshoot. 
This work was supported by the STFC Consolidated Grant ST/L00044X/1.


\end{document}